\documentclass[%
 reprint,
 amsmath,amssymb,
 aps,
pra,
]{revtex4-2}

\usepackage{graphicx}
\usepackage{dcolumn}
\usepackage{bm}
\usepackage{hyperref}
\usepackage{url}
\usepackage{dsfont}
\usepackage[export]{adjustbox}
\usepackage{xcolor}
\usepackage{soul}

\newcommand{\corr}[1]{{#1}}

\begin{document}

\preprint{APS/123-QED}

\title{Wave Topology of Stellar Inertial Oscillations}

\author{Armand Leclerc}
\email{armand.leclerc@ens-lyon.fr}
\author{Guillaume Laibe}
\author{Nicolas Perez}
\affiliation{ENS de Lyon, CRAL UMR5574, Universite Claude Bernard Lyon 1, CNRS, Lyon, F-69007, France}

\date{\today}

\begin{abstract}
Inertial waves in convective regions of stars exhibit topological properties linked to a Chern number of $1$. \corr{The first of these is a unique, unidirectional, prograde oscillation mode within the cavity, which propagates at arbitrarily low frequencies for moderate azimuthal wavenumbers}. The second one are phase singularities around which the phase winds in Fourier space, with winding numbers $\nu = \pm 1$ depending on the hemisphere. \corr{Phase winding is a collective effect over waves propagating in all directions that is strongly robust to noise. This suggests a topology-based method for wave detection in noisy observational data}.
\end{abstract}


\maketitle

\section{Introduction}
Helioseismology has shed unparalleled light on the Sun's interior, and with it stellar interiors in general. While many acoustic modes with mHz frequencies have been identified and used to constrain the rotation and sound speed profiles \cite{basu2016}, other waves are expected to bring complementary information to the surface. Solar internal gravity waves with $10^2\mu$Hz frequencies would bring constraints on the solar core, but confirmation of \corr{their} observation is still missing \cite{garcia2007}. In recent years, a third kind of waves has attracted a great deal of attention. Inertial waves propagate owing to solar rotation with $10^2$nHz frequencies, and are sensitive to the entropy gradient in the convective zone, a quantity to which acoustic modes are essentially insensitive \cite{jones2009}. These waves control the differential rotation of the convective zone and cause it to depart from the Taylor-Proudman columnar structure \cite{bekki2024}. Rossby waves \cite{loptien2018,hathaway2021,gizon2021} and inertial waves of higher frequencies \cite{hanson2022,triana2022} have unequivocally been observed at the solar surface. Inertial modes of convective cores of $\gamma$ Doradus stars have also been observed \cite{ouazzani2020}, as well as Rossby waves \cite{van2016,saio2018,li2019}. Inertial waves are thus an important channel of modern stellar seismology, which motivates further characterization of their oscillation modes.\\
In parallel, recent studies showed a connection between topology and geophysical and astrophysical waves, providing new insight on the properties of global modes \cite{delplace2017,perrot2019,perez2022,leclerc2022,zhu2023,xu2024,lier2024}. In this \corr{study}, we show that stellar inertial waves possess topological properties, characterized by Chern numbers of $\pm1$. This result gives a topological origin to \corr{a unidirectional inertial wave propagating eastward at arbitrarily low frequencies}. It also explains the existence of a phase singularity of the polarization relations of the waves in Fourier space, to which is associated a winding of the phase of $\pm1$. This property provides a novel way of identifying waves. We give a procedure to use this singularity as a signature of the wave and help future identification of seismic signals.

\begin{figure}
    \centering
    \includegraphics[width=\columnwidth]{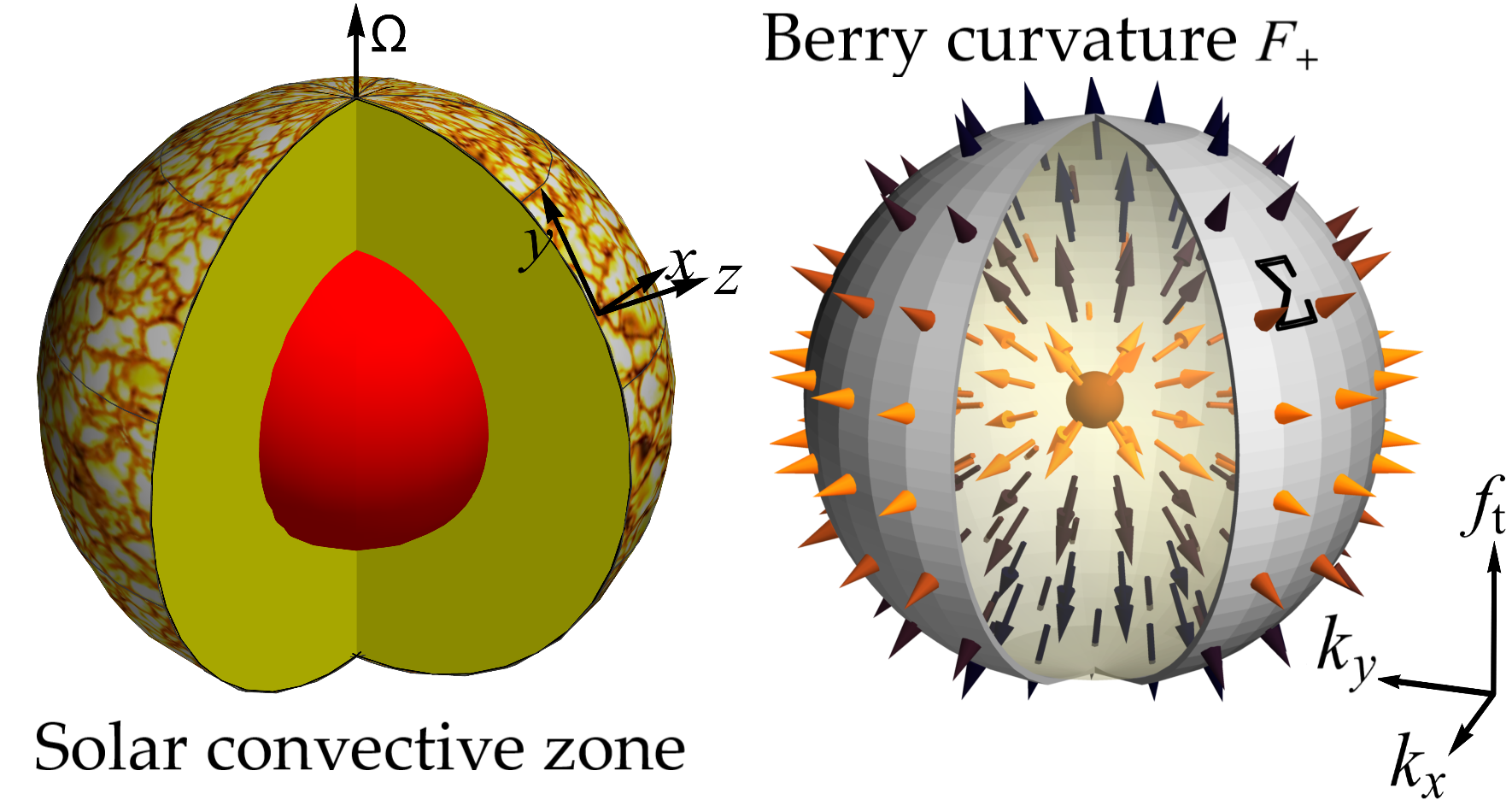}
    \caption{Physical space and parameter space of solar inertial waves. Left: coordinates of the convective zone (yellow). Right: The Berry curvature of the upper inertial waveband is singular at the origin of the parameter space ${k_x=k_y=f_\mathrm{t}=0}$. Brightness indicates the norm of $F_+$.}
    \label{fig:fplane}
\end{figure}

\section{Wave topology of inertial waves}
Consider the rotating convective region of the Sun as represented on Fig.~\ref{fig:fplane}. The medium is radially stratified by the gravity field $-g \mathbf{e}_z$. Following \cite{lockitch1999,ivanov2010}, we assume that the buoyancy frequency ${N^2 = -g\frac{\mathrm{d}\ln\rho_0}{\mathrm{d}z}-\frac{g^2}{c_\mathrm{s}^2}}$ is zero and the sound speed $c_\mathrm{s}$ is uniform. The linearized equations of motions describing fully compressible adiabatic perturbations $(v^\prime,\rho^\prime,p^\prime)$ of this rest state $(\rho_0,p_0)$ are \cite{perez2022thesis}
\begin{eqnarray}
    \partial_t v^\prime &=& -2\Omega\wedge v^\prime -\frac{1}{\rho_0}\nabla p^\prime +\frac{\rho^\prime}{\rho_0^2}\nabla p_0,\label{eq:linear_perturbs1}\\
    \partial_t \rho^\prime &=& -\rho_0\nabla\cdot v^\prime -v^\prime\cdot\nabla\rho_0,\label{eq:linear_perturbs2}\\
    \partial_tp^\prime + v^\prime\cdot\nabla p_0 &=& c_\mathrm{s}^2\left(\partial_t\rho^\prime + v^\prime\cdot\nabla \rho_0\right).\label{eq:linear_perturbs3}
\end{eqnarray}
Velocity components are coupled by the so-called traditional and non-traditional Coriolis parameters ${f_\mathrm{t} = 2 \Omega \sin{\theta}}$ and $f_\mathrm{nt} = 2\Omega\cos{\theta}$ with $\theta$ the latitude. \corr{We follow \cite{papaloizou1978,lockitch1999,vidal2020,jain2023} and retain compressibility in the equations. Compressibility allows to formulate the problem under the form of a linear eigenvalue problem ${\mathcal{L}}X = \omega X$, without bringing any additional complexity to the derivation. Enforcing $\mathrm{div}\, u' =0$ would introduce unnecessary technical difficulties \cite{onuki2024}. Incompressibility is recovered here as a genuine limit of the model.}\\
Essential properties of the inertial waves are captured by an $f$-plane approximation of the rotation, which amounts to assuming that $f_\mathrm{t}$ and $f_\mathrm{nt}$ are constants \cite{vallis2017}. We perform the change of variables $(v,p) = ({\rho_0}^{1/2}v^\prime,{\rho_0}^{-1/2}p^\prime)$ which symmetrizes the equations. Plane-wave solutions $\exp{\left(-i\omega t + ik_x x + ik_y y + ik_zz\right)}$ are then solutions of
\begin{equation}
    \omega \begin{pmatrix}
        v_y\\v_z\\v_x\\p
    \end{pmatrix} = \begin{pmatrix}0&0&-i f_{\rm t}&c_\mathrm{s}k_y\\0&0&if_{\rm nt}& c_\mathrm{s}k_z+iS\\
        if_{\rm t}&-if_{\rm nt}&0&c_\mathrm{s}k_x\\c_\mathrm{s}k_y&c_\mathrm{s}k_z-iS&c_\mathrm{s}k_x&0
    \end{pmatrix}\begin{pmatrix}
        v_y\\v_z\\v_x\\p
    \end{pmatrix},
    \label{eq:fplane-system}
\end{equation}
where $S \equiv \frac{c_\mathrm{s}}{2}\frac{\mathrm{d}\ln\rho_0}{\mathrm{d}z} = -\frac{g}{2c_\mathrm{s}}$ is the stratification parameter, which opens a frequency gap between acoustic and internal \corr{gravity} waves \cite{perrot2019,leclerc2022}.\\
Equation~\eqref{eq:fplane-system} has four eigenvalues, corresponding to two acoustic wavebands and two inertial wavebands. Acoustic waves propagate at frequencies ${\omega^2 \geq \max(4\Omega^2,S^2)}$, whereas inertial waves propagate at frequencies ${\omega^2 \leq 4\Omega^2}$. Hence, stratification also opens a frequency gap between inertial and acoustic waves, as long as $S^2 > 4\Omega^2$, which \corr{is the case for the Solar convective region ($-S \sim 10^3\, \Omega$ from \cite{leclerc2022} and \cite{bekki2024})}. In the following, we shall refer to waves with frequencies $0<\omega_+<2\Omega$ as upper (+) inertial waves and wavebands, and those with frequencies $-2\Omega<\omega_-<0$ as lower (-) inertial waves. Upon inspection of the plane-waves dispersion relation (i.e. the characteristic equation of the matrix in Eq.~\eqref{eq:fplane-system}), the upper and lower inertial waves degenerate at $\omega_+ = \omega_- = 0$ when $k_x=k_y=f_\mathrm{t}=0$, for any $f_\mathrm{nt}$, $k_z$ and $S$. As the degeneracy do not depend on $f_\mathrm{nt}$, $k_z$ and $S$, we treat them as external parameters. \\
\corr{We aim to determine the properties of the waves governed by Eq.~\eqref{eq:fplane-system} that come from topological constraints over the parameter space \cite{delplace2017,perrot2019,perez2022,leclerc2022,zhu2023,xu2024,lier2024}. Let $X_+$ denote the eigenvector of Eq.\eqref{eq:fplane-system} corresponding to the upper inertial waveband. The mathematical space of interest is the fiber bundle $\{{\lambda X_+(k_x,k_y,f_\mathrm{t}) \, | \, \lambda \in \mathds{C}}\}$, which contains all polarization vectors of these waves across the parameter space $(k_x,k_y,f_\mathrm{t})$. The curvature of the fiber bundle, so-called Berry curvature, is}
\begin{equation}
    F_+ \equiv i\tilde{\nabla} \wedge( X_+^\dagger\cdot\tilde{\nabla} X_+),
\end{equation}
\corr{where $\tilde{\nabla} \equiv (\partial_{k_x},\partial_{k_y},\partial_{f_\mathrm{t}/c_\mathrm{s}})^\top$. $F_+$ is a real-valued vector field, which quantifies how the polarization vector $X_+$ is transported when changing the values of parameters $(k_x,k_y,f_\mathrm{t})$ (similarly to Riemannian curvature for transport on Riemannian manifolds).} This vector field is divergence-free \corr{where it is non-singular}, similar to the electrostatic field generated by a point charge. \corr{The Berry curvature is singular whenever the frequency of the wave matches the frequency of another wave of the system, i.e when two eigenvalues of Eq.~\eqref{eq:fplane-system} degenerate. This occurs between the lower and upper inertial waves at $\omega_+ = \omega_- = 0$ for $k_x=k_y=f_\mathrm{t}=0$.} Figure~\ref{fig:fplane} shows $F_+$ around the singular degeneracy point.
\corr{The topology of the manifold of polarization vectors is then characterized by an integer called the Chern number $\mathcal{C}$, which is analogous the Euler characteristic for geometric surfaces. It is defined by}
\begin{equation}
    \mathcal{C} \equiv \frac{1}{2\pi}\oint_\Sigma F\cdot\mathrm{d}\Sigma,
\end{equation}
where $\Sigma$ is any surface enclosing the degeneracy (for example the gray sphere in Fig.~\ref{fig:fplane}). The Chern number is the flux of $F_+$ outward from its singular point. In this problem, the Chern number evaluates to
\begin{equation}
    \mathcal{C}_+ = 1.
\end{equation}
As the sum of Chern numbers over the bands must be zero \cite{delplace2022}, one necessarily has for the lower inertial waveband
\begin{equation}
    \mathcal{C}_- = -1.
\end{equation}

We conclude that stellar inertial waves are {\it topologically non-trivial}, similarly to other waves in geophysical and astrophysical media that where found to have topologically-charged degeneracy points \cite{delplace2017,perrot2019,perez2022,perez2022thesis,leclerc2022}.

\begin{figure}
    \centering
    \includegraphics[width=0.9\columnwidth]{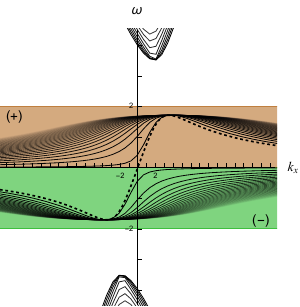}
    \caption{Dispersion relation of the modes in the $\beta$-plane approximation. One mode transits from the lower (-) inertial waveband to the upper one (+), satisfying the spectral flow condition Eq.\eqref{eq:spectral_flow}. Solid lines represent solutions with non-zero $v_y$, while dashed lines indicate solutions with zero $v_y$. Frequencies are shown for $k_z = 1.5$ and $S = -3$, in units of $\Omega$ and $c_\mathrm{s}\Omega^{-1}$.}
    \label{fig:disp_rel_modes_betaplane}
\end{figure}

\section{A spectral flow in inertial standing modes}

The propagation of peculiar edge or interface modes stands out as the manifestation of the existence of non-zero Chern numbers in wave problems \cite{delplace2017,perrot2019,perez2022,leclerc2022,perez2022thesis,qin2023}. They cross frequency gaps between wavebands, and exhibit unique properties such as unidirectionality. These modes exist in virtue of the index theorem \cite{faure2023}, which characterizes the \textit{spectral flow}, a property of the dispersion relation $\{\omega_n(k_x)\}_{n}$ (where $n$ labels the branches) of modes propagating in a medium with varying parameters instead of plane waves in a homogeneous medium. In the case of a single degeneracy point, a waveband $i$ is not composed by the same number of branches in the two limits $k_x\to\pm\infty$. Referring to the difference between these branch numbers as $\Delta\mathcal{N}_i$, the index theorem states that $\Delta\mathcal{N}_i = \mathcal{C}_i$: the Chern number gives the number of modes leaving/arriving into the waveband when $k_x$ increases. For the inertial waves,
\begin{equation}
    \Delta\mathcal{N}_+ = -\Delta\mathcal{N}_- = +1.\label{eq:spectral_flow}
\end{equation}
Equation~\ref{eq:spectral_flow} guarantees the presence of a mode of topological origin transiting between the (-) and the (+) bands, similarly to what has been found in other studies (e.g. Fig.10 of \cite{bekki2022}, or Fig.5 of \cite{jain2023}). This mode is the \corr{stable counterpart of Busse columns, the most unstable mode of a convectively unstable rotating fluid \cite{busse1970}. It is sometimes called the Busse mode. It is stable here as we study neutral stratification.} This wave has been seen in numerical works \cite{glatzmaier1981,bekki2022}, as well as in the analytical study of \cite{jain2023} which models the equator as a channel with $f_\mathrm{t}=0$ and impenetrable boundaries. It also has been identified recently in a DNS of nonlinear inertial waves in the Sun \cite{blume2024}. The value of the Chern number provides an explanation as to why it is purely prograde. \\
We confirm the existence and the properties of this topological mode by finding the normal modes of Eqs.~\eqref{eq:linear_perturbs1}-\eqref{eq:linear_perturbs3} under the $\beta$-plane approximation. The latter consists in expanding the spatial dependence of $f_\mathrm{t}$ and $f_\mathrm{nt}$ at first order in latitude $y$, which can be expressed as
\begin{eqnarray}
    f_\mathrm{t} = \beta y,\\
    f_\mathrm{nt} = 2\Omega.
\end{eqnarray}
The plane-parallel geometry of the $\beta$-plane retains the frequency behavior of the modes, but does not capture the columnar structure found in shell cavities \cite{busse1970,glatzmaier1981}. \\
The equations of normal modes associated to linear perturbations of the form $\exp{\left(-i\omega t + ik_x x + ik_zz\right)}\begin{pmatrix} v_x(y) & v_y(y) & v_z(y) & p(y) \end{pmatrix}^\top$ are
\begin{equation}
    \omega \begin{pmatrix}
        v_y\\v_z\\v_x\\p
    \end{pmatrix} = \begin{pmatrix}0&0&-i\beta y&-ic_\mathrm{s}\partial_y\\0&0&i2\Omega& c_\mathrm{s}k_z+iS\\
        i\beta y&-i2\Omega&0&c_\mathrm{s}k_x\\-ic_\mathrm{s}\partial_y&c_\mathrm{s}k_z-iS&c_\mathrm{s}k_x&0
    \end{pmatrix}\begin{pmatrix}
        v_y\\v_z\\v_x\\p
    \end{pmatrix}.
    \label{eq:betaModel}
\end{equation}
Boundaries in the vertical direction discretize $k_z$. Imposing impenetrable boundary conditions $v_z = 0$ yields $k_z = \frac{p\pi}{L}$, where $L$ is the width of the convective region and $p$ is a positive integer.\\
Equation~\eqref{eq:betaModel} can be reduced to a single ordinary differential equation on ${\phi \equiv \exp\left(\frac{4i\beta \Omega c_\mathrm{s}k_z}{\omega^2-4\Omega^2}\frac{y^2}{2}\right)v_y}$ (see Appendices):
\begin{equation}
    -\partial_{yy}\phi + \left(\gamma y^2 +\delta\right)\phi = 0,\label{eq:almostQHO}
\end{equation}
where $\gamma = \frac{\beta^2}{\omega^2-4\Omega^2}(\omega^2 - c_\mathrm{s}^2k_z^2 - S^2 - \frac{4c_\mathrm{s}^2k_z^2\Omega^2}{\omega^2-4\Omega^2})$, ${\delta = \frac{1}{\omega^2-4\Omega^2}(\beta(c_\mathrm{s}k_x\omega+2S\Omega)+A\omega)}$ and
\begin{equation}
    A \equiv \det \begin{pmatrix}
        -\omega&i2\Omega& c_\mathrm{s}k_z+iS\\
        -i2\Omega&-\omega&c_\mathrm{s}k_x\\c_\mathrm{s}k_z-iS&c_\mathrm{s}k_x&-\omega
    \end{pmatrix}.
\end{equation}
For $\gamma < 0$, Eq.~\eqref{eq:almostQHO} \corr{has no non-zero square-integrable solution}. For $\gamma > 0$, Eq.\eqref{eq:almostQHO} is transformed into a Weber differential equation \cite{whittaker2021} through the change of coordinate $\xi = \sqrt{2\gamma}y$
\begin{equation}
    -\partial_{\xi\xi}\phi + \left(\frac{\xi^2}{4}-\frac{1}{2}-(\frac{-\delta}{2\gamma}-\frac{1}{2})\right)\phi=0.
\end{equation}
Regularity of $\phi$ at infinity provides a condition for the normal modes in the $\beta$-plane model, imposing
\begin{equation}
    \frac{-\delta}{2\gamma}-\frac{1}{2} = n-1,\label{eq:global_dispRel_vy}
\end{equation}
for $n\geq1$ any positive integer. Combined with the condition $\gamma > 0$, Eq.~\eqref{eq:global_dispRel_vy} provides the dispersion relation for the normal modes. The typical latitudinal extent of the waves is $\gamma^{-1/2}$. The eigenfunctions of the modes are given by Hermite polynomials $H_n$ with
\begin{equation}
    \phi_n(\xi) = \exp{\left(-\frac{\xi^2}{4}\right)}H_{n-1}(\frac{\xi}{\sqrt{2}}).
\end{equation}


\begin{figure}
    \centering
    \includegraphics[width=\columnwidth]{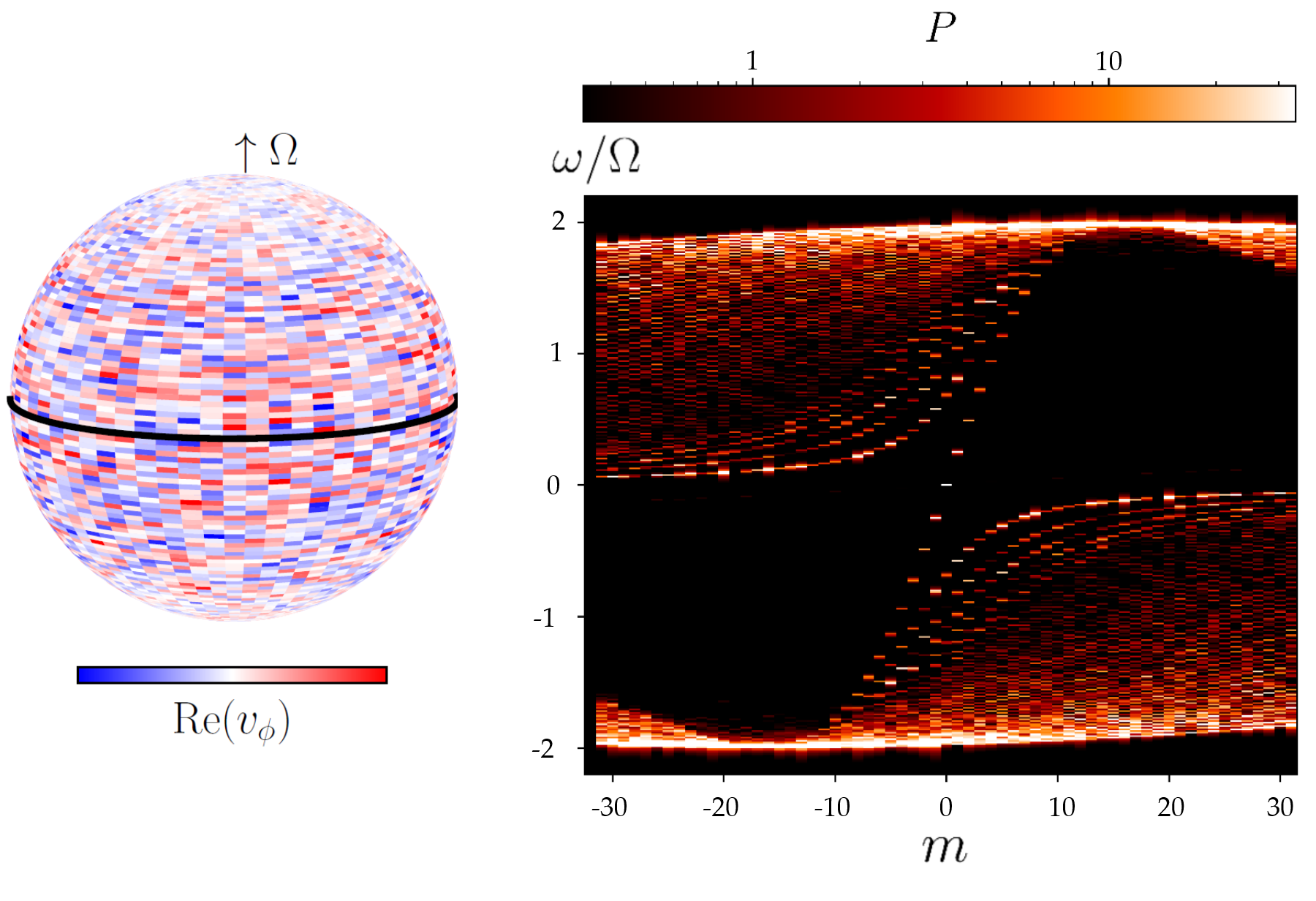}
    \caption{A branch of modes transits between the inertial wavebands in numerical simulations of linear inertial waves on a sphere. Left: Snapshot of zonal velocity. Right: Power spectrum at the equator. Ridges, i.e. the bright regions of the power spectrum, are the normal modes. \corr{Power is measured in arbitrary units (equations are linear).}}
    \label{fig:sim-results}
\end{figure}

These solutions are the normal modes that have non-zero latitudinal velocity $v_y$. To identify all the normal modes, one must also determine those with zero $v_y$, denoted as $n=0$. From Eq.\eqref{eq:betaModel}, those satisfy
\begin{equation}
    A = 0 = -\omega^3 +(4\Omega^2+S^2+c_\mathrm{s}^2k_x^2+c_\mathrm{s}^2k_z^2)\omega + 4\Omega Sc_\mathrm{s}k_x.\label{eq:dipersion_rel_vyzero}
\end{equation}
Three modes are solutions of this equation: one inertial wave, and two acoustic waves. Analytical expressions of their dispersion relations are given in Appendices.\\
These three modes complete the spectrum given by Eq.\eqref{eq:global_dispRel_vy}, for any radial order $p$ and latitudinal order $n$. Figure~\ref{fig:disp_rel_modes_betaplane} shows the dispersion relations obtained for all modes. It appears clearly that the $n=0$ inertial branch transits from the lower inertial band to the upper one as $k_x$ increases. This is the spectral flow expected from the Chern numbers $\mathcal{C}=\pm1$ of the wavebands. This mode possesses the distinctive characteristic of being unidirectional, exclusively prograde, while the other modes can propagate either eastward or westward.\\
\corr{Fig.~\ref{fig:sim-results} confirms the presence of this prograde wave among the oscillation modes of spherical shells in numerical simulations}.

\section{Phase singularity and winding}
\begin{figure}
    \centering
    \includegraphics[width=\columnwidth]{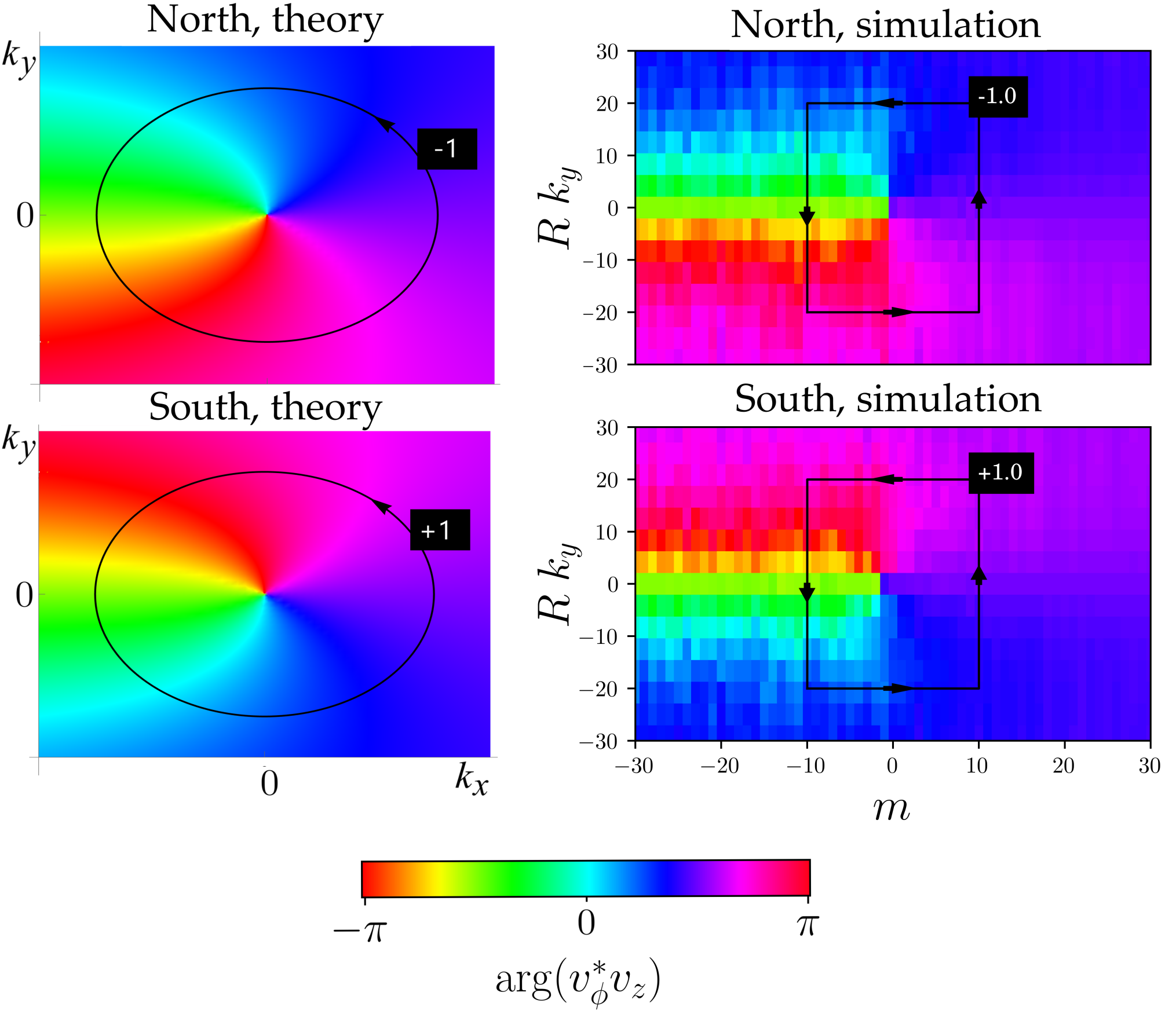}
    \caption{Phase singularity of the upper (+) inertial waves. Left: phase in the $f$-plane. Right: phase found when analyzing the simulations Northern (Southern) hemisphere only.}
    \label{fig:phase_singularity}
\end{figure}

Beyond evidencing the spectral flow, we will now show a way to exploit another topological property of the inertial modes: a physical quantity known as phase winding, measuring the cumulative phase difference between two complex quantities over a closed path in the parameter space, which attracted interest in recent years, e.g. in geophysical waves \cite{zhu2023,xu2024} or material sciences \cite{ergoktas2024}. Non-zero Chern numbers imply the existence of phase singularities (see Appendices, or \cite{fosel2017} for an example in condensed matter). The relative phase between two wave components exhibits a singular point in Fourier space, around which the phase winds. Phase winding changes sign when changing hemisphere. Figure~\ref{fig:phase_singularity} (left) shows that the relative phase between the zonal and vertical velocity perturbations $v_x$ and $v_z$ of plane-waves in an $f$-plane, i.e the components of the eigenvector of Eq.~\eqref{eq:fplane-system}, is singular the origin of the $(k_x,k_y)$ plane. Formally, the winding number of this phase is
\begin{equation}
    \nu = \frac{1}{2\pi}\oint_\Gamma \Tilde{\nabla}\mathrm{arg}(v_x^*v_z) \cdot \mathrm{d}k,
\end{equation}
where $\Gamma$ can be any closed loop encircling the origin once with counterclockwise orientation. Depending on the upper (+) or lower (-) inertial waveband and the hemisphere, one finds
\begin{eqnarray}
    \nu_{+,\;{\rm north}} &=& -1,\label{eq:windings_first}\\
    \nu_{+,\;{\rm south}} &=& +1,\\
    \nu_{-,\;{\rm north}} &=& +1,\\
    \nu_{-,\;{\rm south}} &=& -1.\label{eq:windings_last}
\end{eqnarray}

Physically, this value of $\nu$ means that the peaks of $v_z$ have precisely shifted by one wavelength with respect to the peaks of $v_x$ after completing one full revolution along the contour. In contrast, the relative phases between $v_x$ and $v_y$, or $v_x$ and $p$, do not experience winding and remain non singular.\\
\corr{Using numerical simulations, we confirm that the phase singularity identified in the $f$-plane persists in spherical geometry. This aligns with the results of \cite{xu2024}, who observed phase winding of Poincaré waves in the atmospheric data of the Earth, consistent with $f$-plane predictions. For this purpose, we solve Eqs.~\eqref{eq:linear_perturbs1}-\eqref{eq:linear_perturbs3} within the solar convective zone. For simplicity, we fix the vertical wavenumber to $k_z = p\pi/L$, restricting the dynamics to a specific radial order $p$. We compute the two-dimensional dynamics in $(\theta,\phi)$ for modes with radial order $p=1$ with the code \textsc{dedalus} \cite{burns2020}.}
We use $\Omega^{-1}$ and $c_\mathrm{s}\Omega^{-1}$ as units of time and length. The evolution equations of linear perturbations become 
\begin{eqnarray}
    \partial_t u + 2\mathbf{e}_r\wedge u + \nabla p + 2\sin(\theta) v_z \mathbf{e}_\phi &=& 0,\label{eq:eq_on_sphere1}\\
    \partial_t v_z +(ik_z- S)p -2\sin(\theta) u \cdot \mathbf{e}_\phi &=& 0,\label{eq:eq_on_sphere2}\\
    \partial_t p + \nabla \cdot u +(ik_z+ S)v_z &=&0,\label{eq:eq_on_sphere3}
\end{eqnarray}
where $u = (u_\theta,u_\phi) = (-v_y,v_x)$ are the angular components of the velocity in spherical coordinates, and $\nabla$ is the nabla operator on the sphere. The evolution equations are written in a covariant way for the angular directions, a necessity for the spectral decomposition by \textsc{dedalus}. (See Appendices for details of implementation).\\
The fields are initialized with random values across the grid to excite all wavelengths. After solving for the evolution, the velocity data $u(t,\theta,\phi)$ and $v_z(t,\theta,\phi)$ are then Fourier-transformed in both space and time. For consistency, we adopt the convention
\begin{equation}
    \hat{f}(\omega,k_y,m) = \int f(t,\theta,\phi)\mathrm{e}^{-i\omega t - ik_y R \theta +im\phi}\mathrm{d}t\mathrm{d}\theta\mathrm{d}\phi.
\end{equation}
This transform is computed numerically for all values of $\phi$ and $t$, but on various \corr{domains} of latitude $y=R(\frac{\pi}{2}-\theta)$. The azimuthal wavenumber $m$ is equated to $R k_x$.\\
We compute the power spectrum of the kinetic energy $P = \vert\hat{v_x}\vert^2+\vert\hat{v_y}\vert^2+\vert\hat{v_z}\vert^2$ at the equator ($y = 0$) so as to extract equatorial waves. This yields the dispersion relation of the modes in an $(m,\omega)$ diagram. Figure~\ref{fig:sim-results} (right panel) shows this power spectrum; ridges of power indicate normal modes which are in agreement with the propagation of the topological mode found in the $\beta$-plane. \\
In order to extract the phase singularities, we compute Fourier transforms of the data over an entire hemisphere exclusively, which yields the relative phase $\arg(\hat{u_\phi}^*\,\hat{v_z})$$(\omega,m,k_y)$. This quantity is subsequently averaged on the frequencies of the upper inertial band $0<\omega<2\Omega$. Results are shown on Fig.~\ref{fig:phase_singularity}(right), which reveals the phase singularity discussed above without any ambiguity. The phase winding values correspond to those of Eqs.~\eqref{eq:windings_first}-\eqref{eq:windings_last}.\\
We repeat the aforementioned analysis, adding various noises with intensities up to approximately 10 times higher than that of the initial signal (see Appendices for details). \corr{Figure~\ref{fig:noise_robustness} shows the relative phase in case of noisy data, for which the measurement of $\nu$ still yields $-1$ in the Northern hemisphere.} The determination of the winding number is robust to a significant level of noise, a common characteristic of topological properties in physics \cite{delplace2022}.

\begin{figure}
    \centering
    \includegraphics[width=\columnwidth]{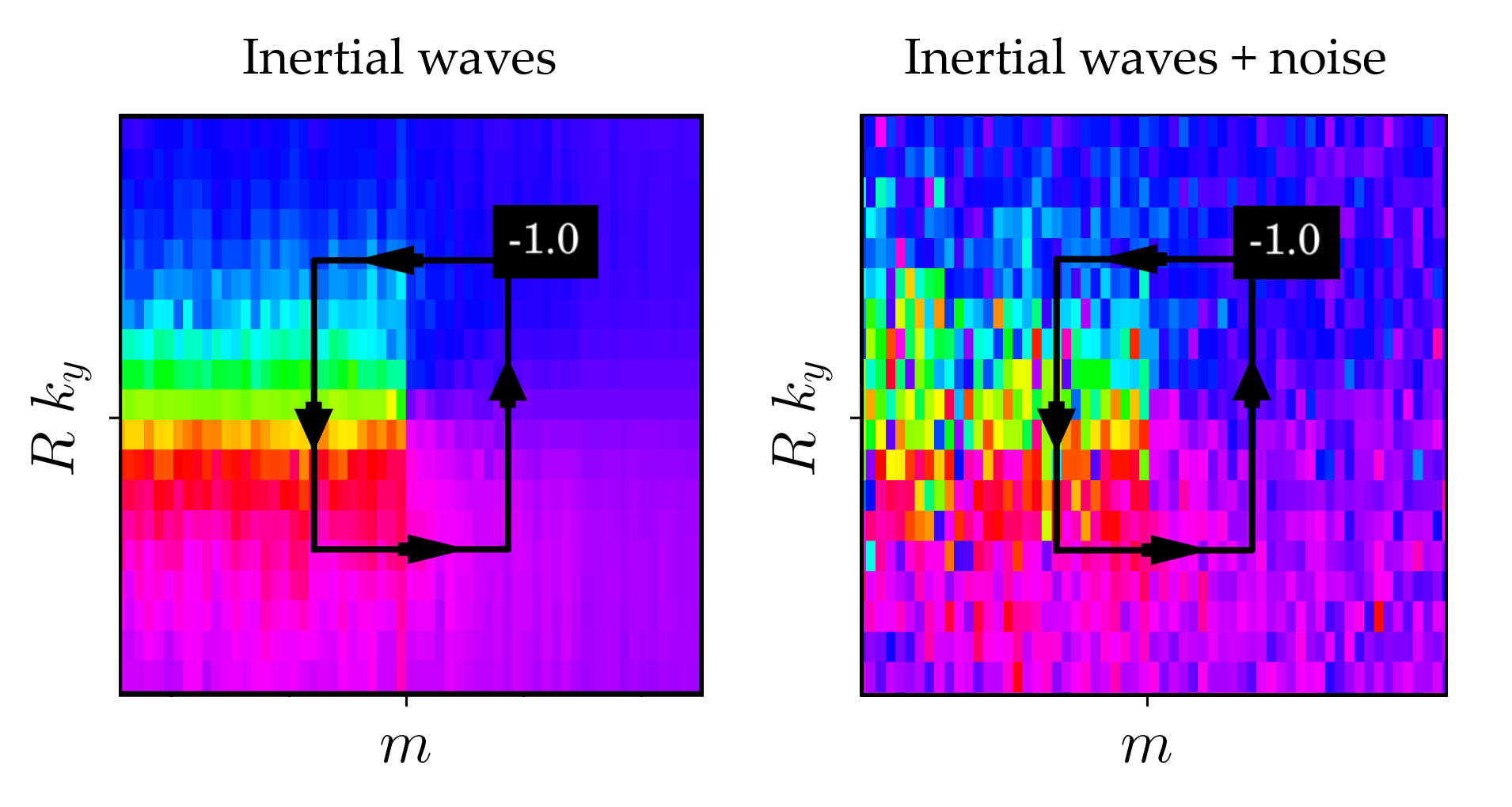}
    \caption{The winding of the phase is robust to the addition of noise in the data. On a signal of pure inertial waves (left), or upon adding a random noise on the velocities data (right), the measurement of the winding $\nu$ still yields $-1$. The amplitude of the noise is 10 times that of the initial condition of the simulation.}
    \label{fig:noise_robustness}
\end{figure}

\section{Conclusion}

Inertial waves in convective stellar regions have topological properties linked to a parameter space degeneracy at the equator and a Chern number of $\pm$1. We derived this result \corr{accounting for compressibility for both generality and simplicity, although compressible effects are not important in this problem, since frequencies of inertial and acoustic waves are of distinct orders of magnitude in the Sun.}

\corr{Wave topology predicts the existence of a mode of spectral flow between the two $(+)$ and $(-)$ inertial bands. This mode uniquely propagates at arbitrarily low frequencies $\omega$ for moderate azimuthal wavelengths (Fig.~\ref{fig:disp_rel_modes_betaplane}) and as such, may be significant for certain stellar objects. For instance, it is expected to propagate at arbitrarily low frequencies within the convective cores of $\gamma$ Dor stars, while gravity-inertial waves propagate in the outer layers. Because of their low frequencies, high-order $g$-modes are likely to couple with the core mode, making it a valuable probe for core dynamics \cite{ouazzani2020}. This also implies that mixed modes should consistently appear in the spectra of $\gamma$ Dor stars.}

\corr{Wave topology further predicts a collective phase winding of $\pm 1$ for sets of inertial waves propagating in multiple directions. Phase winding is calculated by accumulating the phase of the waves as their propagation direction varies along a closed loop in parameter space. Consequently, measuring phase winding may be more feasible using observational data than detecting individual waves, as it integrates the power spectrum across parts of the wavebands. This approach allows for observational diagnosis even when the signal is weak.}

\corr{Topology ensures finally that the conclusions of this study apply to any cavity containing an equator, whether it involves a fully convective star or a convective shell zone. The rotation profile can be arbitrary as long as it is Rayleigh stable. Topological tools used in Hermitian physics are however} not suited for discussing viscous damping (Ekman layer). Analyzing its effects requires a non-Hermitian topological approach \cite{jezequel2023,zhu2023}. Fortunately, viscous damping in the Sun occurs on a much longer timescale than rotation \cite{canuto1998}, and \corr{may only affect the results of this study as small corrections}.

\begin{acknowledgments}
We acknowledge funding from the ERC CoG project
PODCAST No 864965. AL is funded by Contrat Doctoral Spécifique Normalien.
\end{acknowledgments}

\newpage 
\appendix

\section{Analytical derivation of modes}
\label{app:analyt_beta_modes}
The system of equations Eq.~\eqref{eq:betaModel} governing the evolution of normal modes in the $\beta$-plane can be written under the form
\begin{eqnarray}
    \omega v_y +if_\mathrm{t}v_x + ic_\mathrm{s}\partial_yp &=& 0,\label{eq:first_beta_plane}\\[6pt]
    \begin{pmatrix}
        -\omega & if_\mathrm{nt} & c_\mathrm{s}k_z + iS \\
        -if_\mathrm{nt} &  -\omega & c_\mathrm{s}k_x      \\
        c_\mathrm{s}k_z - iS & c_\mathrm{s}k_x& -\omega 
    \end{pmatrix}\begin{pmatrix}
        v_z \\ v_x \\ p
    \end{pmatrix} &=& \begin{pmatrix}
        0 \\ -if_\mathrm{t}v_y \\ ic_\mathrm{s}\partial_yv_y
    \end{pmatrix}.\hspace{0.8cm}\label{eq:matrix_beta_plane}
\end{eqnarray}

The matrix in the left-hand side of Eq.~\eqref{eq:matrix_beta_plane} involves no $y$-derivative, which allows us to express $v_z$, $v_x$ and $p$ as linear combinations of $v_y$ and $\partial_yv_y$ by inverting it. Using the result in Eq.~\eqref{eq:first_beta_plane} yields to an ordinary differential equation on $v_y$. In details, define
\begin{equation}
    A \equiv \det \begin{pmatrix}
        -\omega&i2\Omega& c_\mathrm{s}k_z+iS\\
        -i2\Omega&-\omega&c_\mathrm{s}k_x\\c_\mathrm{s}k_z-iS&c_\mathrm{s}k_x&-\omega
    \end{pmatrix},
\end{equation}
such that $A\neq 0$ when $v_y$ is a non-zero function. Hence, 
\begin{widetext}
\begin{eqnarray}
    v_z &=& \frac{1}{A}\bigg[\beta y\bigg(c_\mathrm{s}k_x(-ic_\mathrm{s}k_z+S)+2\Omega\omega\bigg)v_y +  \label{eq:isolated_vz_betamodel}\bigg((ic_\mathrm{s}k_z-S)\omega - 2 c_\mathrm{s}k_x \Omega\bigg)c_\mathrm{s}\partial_y v_y\bigg],\\
    v_x &=& \frac{1}{A}\bigg[i\beta y \bigg(S^2+c_\mathrm{s}^2k_z^2-\omega^2\bigg)v_y + \label{eq:isolated_vx_betamodel}i\bigg(c_\mathrm{s}k_x\omega-(ic_\mathrm{s}k_z-S)2\Omega\bigg)c_\mathrm{s}\partial_y v_y\bigg],\\
    p &=& \frac{1}{A}\bigg[-i\beta y\bigg(c_\mathrm{s}k_x\omega+(ic_\mathrm{s}k_z+S)2\Omega\bigg)v_y +\label{eq:isolated_p_betamodel}i\bigg(\omega^2-4\Omega^2\bigg)c_\mathrm{s}\partial_y v_y\bigg].
\end{eqnarray}

Casting Eqs.\eqref{eq:isolated_vx_betamodel} and \eqref{eq:isolated_p_betamodel} into Eq.\eqref{eq:first_beta_plane} yields

\begin{eqnarray}
    &&(\omega^2-4\Omega^2)c_\mathrm{s}^2\partial_{yy}v_y - 4i\beta \Omega c_\mathrm{s}^2k_z\: y\;\partial_y v_y +\label{eq:ODEvy}\bigg(\beta^2(c_\mathrm{s}^2k_z^2+S^2-\omega^2)y^2-\beta(c_\mathrm{s}k_x\omega+2\Omega(ic_\mathrm{s}k_z+S))-A\omega\bigg)v_y = 0.
\end{eqnarray}
\end{widetext}

The change of variable $\phi \equiv \exp\left(-\frac{i\beta \Omega k_z}{\omega^2-4\Omega^2}y^2\right)v_y$ removes the second term of the left-hand side of Eq.\eqref{eq:ODEvy}, which then reduces to
\begin{equation}
    -\partial_{yy}\phi + \left(\gamma y^2 +\delta\right)\phi = 0,\label{eq:almostQHO_app}
\end{equation}
with 
\begin{eqnarray}
    \gamma &=& -\frac{\beta^2}{\omega^2-4\Omega^2}(c_\mathrm{s}^2k_z^2+S^2-\omega^2+\frac{4c_\mathrm{s}^2k_z^2\Omega^2}{\omega^2-4\Omega^2}),\\
    \delta &=& \frac{1}{\omega^2-4\Omega^2}(c_\mathrm{s}\beta(c_\mathrm{s}k_x\omega+2S\Omega)+A\omega),
\end{eqnarray}
yielding Eq.~\eqref{eq:almostQHO} discussed in the main text.\\

From Eq.~\eqref{eq:first_beta_plane}, the modes with $v_y = 0$ must satisfy $A=0$. The three roots of this third order polynomial describe one inertial wave ($-2\Omega\leq\omega\leq2\Omega$) and two acoustic waves ({$\omega^2 \geq S^2$}). Following the method of \cite{oldham2009}, we obtain the expressions of the three roots. The one corresponding to the inertial mode is
\begin{eqnarray}
    &&\omega_{\mathrm{inertial},n=0} = \frac{2}{\sqrt{3}}\sqrt{c_\mathrm{s}^2(k_x^2+k_z^2)+S^2+4 \Omega ^2}\times\hspace{1cm}\\
    &&\hspace{1cm} \cos \left(\frac{\arccos\left(\frac{6 \sqrt{3} c_\mathrm{s}k_x S \Omega }{\sqrt{\left(c_\mathrm{s}^2(k_x^2+k_z^2)+S^2+4 \Omega ^2\right)^3}}\right) - 2\pi }{3}\right).\nonumber
\end{eqnarray} 
The two $n=0$ acoustic modes have frequencies
\begin{eqnarray}
    &&\omega_{\mathrm{acoustic},n=0} = \frac{2}{\sqrt{3}}\sqrt{c_\mathrm{s}^2(k_x^2+k_z^2)+S^2+4 \Omega ^2}\times\hspace{1cm}\\
    &&\hspace{1cm} \cos \left(\frac{\arccos\left(\frac{6 \sqrt{3}c_\mathrm{s} k_x S \Omega }{\sqrt{\left(c_\mathrm{s}^2(k_x^2+k_z^2)+S^2+4 \Omega ^2\right)^3}}\right) + 2 s \pi }{3}\right),\nonumber
\end{eqnarray}
where the case $s=0$ and $s=+1$ corresponds to positive and negative frequencies respectively.

\section{Non-zero Chern numbers and phase winding}
Under the appropriate gauge choice $\phi_{a}$, a normalised eigenvector of the upper inertial waveband can be written
\begin{eqnarray}
\Psi &\equiv& \begin{pmatrix}
        a \\ b\: \mathrm{e}^{i\phi_b}\\ c\:\mathrm{e}^{i\phi_c}\\d\:\mathrm{e}^{i\phi_d}
    \end{pmatrix},  \label{eq:eigenvector} \\
    &=& \mathrm{e}^{-i \phi_{ a}}\begin{pmatrix}
    i f_\mathrm{t} k_x +\frac{2\Omega}{\omega} f_\mathrm{t} k_z  +  i \frac{2\Omega}{\omega}f_\mathrm{t} S  -k_y \omega + \frac{4\Omega^2}{\omega} k_y \\
    \frac{2 \Omega}{\omega}  (f_\mathrm{t} k_y-i k_x \omega )+ \frac{f_\mathrm{t}^2-\omega^2}{\omega}(k_z+i S)\\-i f_\mathrm{t} k_y-k_x \omega +2 i k_z \Omega -2 S \Omega \\f_\mathrm{t}^2-\omega ^2+4 \Omega ^2
\end{pmatrix},\nonumber
\end{eqnarray}
where $a,b,c,d$ are real numbers satisfying $a^2+b^2+c^2+d^2 = 1$. The Chern number of the band is the gauge-invariant quantity
\begin{equation}
    \mathcal{C} = \frac{1}{2\pi}\oint i \nabla \times (\Psi^\dagger\cdot\nabla\Psi)\cdot\mathrm{d}\Sigma,
    \label{eq:ChernAppendix}
\end{equation}
where $\Sigma$ is a closed oriented surface enclosing the degeneracy point located at the origin of parameter space $(k_x,k_y,f_\mathrm{t}) = (0,0,0)$. With the functional form given by Eq.~\eqref{eq:eigenvector}, one obtains
\begin{widetext}
\begin{eqnarray}
    2\pi\mathcal{C} &=& \oint_{\Sigma} i\nabla \wedge (\Psi^\dagger\cdot\nabla\Psi)\cdot\mathrm{d}\Sigma,\\
    &=& i\oint_{\Sigma} \nabla \wedge \left(a\nabla a + b\nabla b + c\nabla c + d\nabla d + i(b^2\nabla\phi_b+c^2\nabla\phi_c+d^2\nabla\phi_d)\right)\mathrm{d}\Sigma,\\
    &=& \frac{i}{2}\oint_{\Sigma} \nabla \wedge \underbrace{\nabla\left(a^2 + b^2 + c^2 +d^2\right)}_{=0} -\oint_{\Sigma} \nabla\wedge(b^2\nabla\phi_b+c^2\nabla\phi_c+d^2\nabla\phi_d)\mathrm{d}\Sigma, \\
    &=&-\oint_{\Sigma} \nabla\wedge R \, \mathrm{d}\Sigma\label{eq:integralAsRot} ,
\end{eqnarray}
\end{widetext}
where $R \equiv b^2\nabla\phi_b+c^2\nabla\phi_c+d^2\nabla\phi_d$. Hence, if $R$ is smooth everywhere on $\Sigma$, $\mathcal{C} = -\frac{1}{2\pi} \int_{\partial \Sigma} R \;\mathrm{d}p = 0$ from Stokes theorem. Conversely,  a non-zero Chern number implies for $R$ to be singular, i.e the existence of phase singularity in the parameter space.
%
%

\section{Simulations}
\label{app:sims}
\begin{figure}
    \centering
    \includegraphics[width=1\columnwidth]{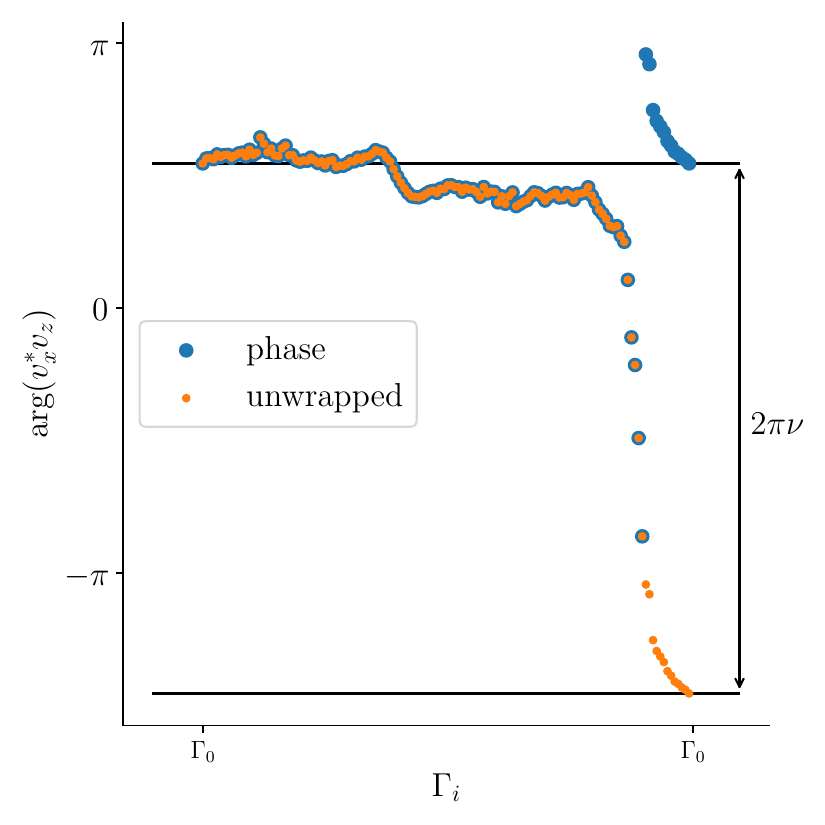}
    \caption{The winding number of the closed loop $\Gamma$ is measured by unwraping the phase. A continuous curve is obtained, and the difference between the first and last point is $2\pi\nu$.}
    \label{fig:unwraping}
\end{figure}

The equations of linear inertial waves on the sphere Eqs.~\eqref{eq:eq_on_sphere1}-\ref{eq:eq_on_sphere3} are solved using the spectral code \textsc{dedalus} \cite{burns2020}. The spatial solver decomposes the fields on bases of polynomials in angular coordinates $(\theta,\phi)$. The bases are truncated at order $(N_\theta,N_\phi) = (128,64)$. The timestepping solver is a Runge-Kutta method of order 2, with constant timestep $\mathrm{d}t = 10^{-2} \Omega^{-1}$. The four fields $u_\theta,u_\phi,v_z,p$ are initiated with uniformly sampled random values between -1 and 1 on the $128\times64$ grid points. The evolution is then solved until $t = 200 \;\Omega^{-1}$. In these units, the solar radius is $R = 2.0\;10^{-2}\;c_\mathrm{s}\Omega^{-1}$. We set $S = -800 \Omega$ (values discussed e.g. in \cite{leclerc2022} and \cite{bekki2024}).\\
The Fourier transforms of the output of the simulations are computed by the Fast Fourier Transforms (FFT) methods of Numpy. The spatial Fourier transform in the Northern is computed on all the points $0<y/R<\pi/2$.  The Fourier transform in the Southern hemisphere is computed on all the points $0>y/R>-\pi/2$.\\
The winding number is measured by unwrapping the signal of the phase along the closed path (black loop on Fig.~\ref{fig:phase_singularity}). Unwrapping the signal involves adjusting large jumps by multiples of the period $2\pi$ to obtain a continuous function that is not restricted to $[-\pi,\pi]$. The winding number is then given by the difference between the first and last values of the signal, which is an integer multiple of $2\pi$. Figure~\ref{fig:unwraping} shows the raw signal of the phase along the points $\Gamma_i$ of the loop $\Gamma$, and the result of its unwrapping. This procedure is guaranteed to yield an integer result for $\nu$, ensuring that any error would be at least 100\%.\\
The full Python script used for this study is available at \href{https://www.github.com/ArmandLeclerc/topo-inertial-waves}{https://www.github.com/ArmandLeclerc/topo-inertial-waves}.

\section{Interpretation of phase winding}
\begin{figure}
    \centering
    \includegraphics[width=1\columnwidth]{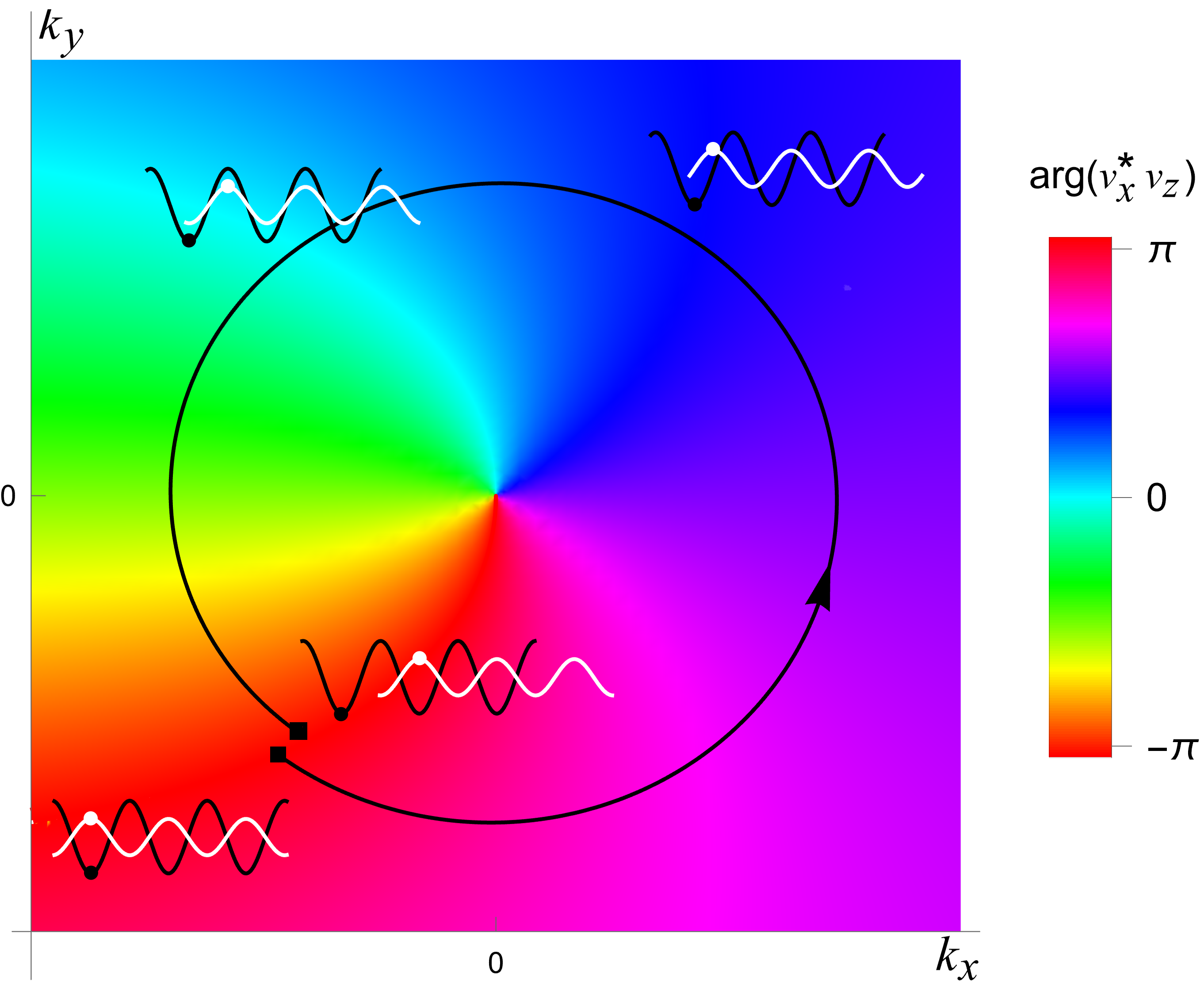}
    \caption{The phase winding of 1 corresponds to a $2\pi$ shift of the phase when completing a closed path around the topological singularity in the Fourier space.}
    \label{fig:windin_schema}
\end{figure}
When the eigenvector associated with the inertial wave in Fourier space is continuously varied, the relative phase between $v_x$ and $v_z$ changes. Starting and coming back to given point on a closed path in the parameter space, the resulting plane wave is physically identical to the initial one. This requirement implies the phase accumulated along the closed path is not identically zero, but a multiple of $2 \pi$. Figure~\ref{fig:windin_schema} shows phase evolution along the path that corresponds to a winding of 1 for the inertial wave. The crests of $v_z$ are shifted exactly by one wavelength with respect to the crests of $v_x$ when moving along the black loop that encloses the singularity.

\section{Robustness to noise}
\label{app:noise}

\begin{figure}
    \centering
    \includegraphics[width=1\columnwidth]{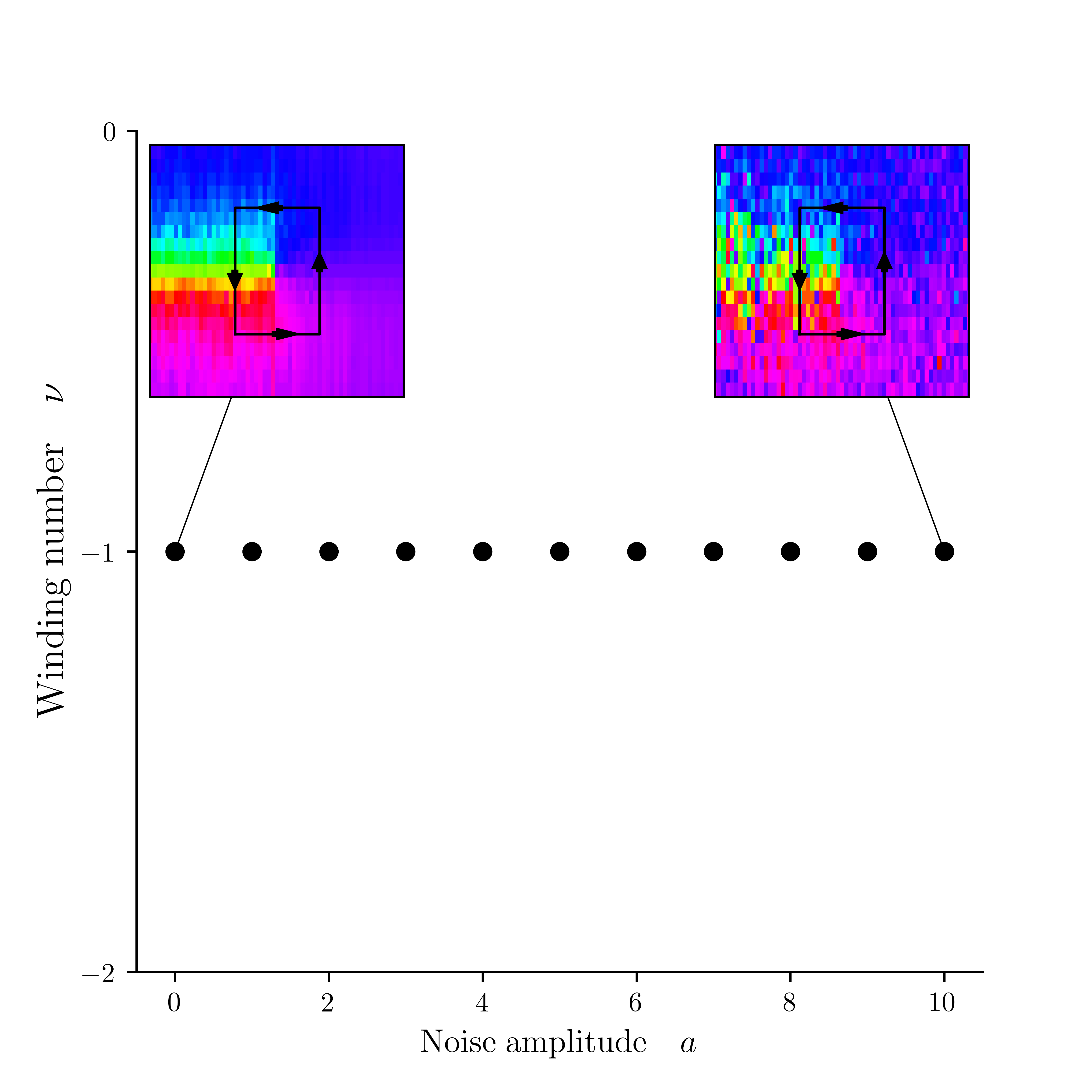}
    \caption{Winding numbers measured as a function of the relative amplitude of noise with respect to the initial linear perturbation. Up to a ratio of $a = 10$, winding numbers are still measured to be $-1$.}
    \label{fig:app_noise}
\end{figure}

\begin{figure}
    \centering
    \includegraphics[width=1\columnwidth]{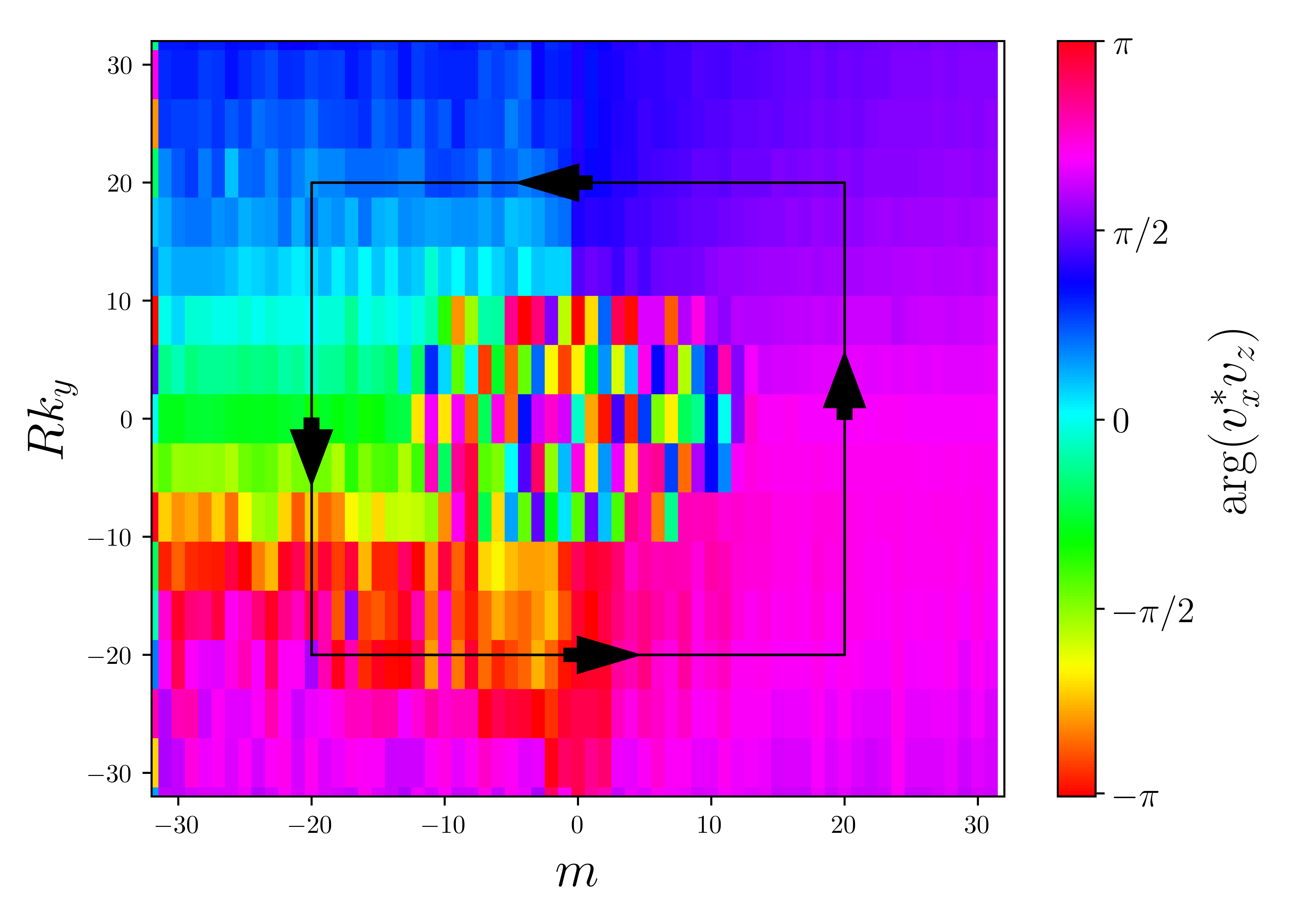}
    \caption{Example of phase winding measurement with isotropic noise whose power spectrum decreases at high spatial frequencies.}
    \label{fig:gaussian_noise}
\end{figure}

An artificial amount of noise is added to the velocity data calculated by the simulation by
\begin{eqnarray}
    u_\mathrm{tot} &=& u_\mathrm{sim } \hphantom{_{z}} +  u_\mathrm{noise},\\
    v_\mathrm{z,tot} &=& v_{z,\mathrm{sim}} + v_{z,\mathrm{noise}},
\end{eqnarray}
where the noise fields are generated by sampling random values between $-a$ and $a$ on the 128$\times$64 grid points at every time step. $a$ is the noise amplitude relative to the amplitude of the initial amplitude of the linear perturbation. The winding number $\nu_+$ in the northern hemisphere is then measured, and shown on Fig.\ref{fig:app_noise}. Up to a noise factor $a = 10$, the winding number is still measured to be $-1$, demonstrating the robustness of the method against noise. Physical noise sources can produce noise with a structured power spectrum, such as Gaussian noise or power-law noise. Figure~\ref{fig:gaussian_noise} shows that by employing a contour of integration where the noise has significantly diminished relative to the wave signal, it is possible to measure the phase winding.


\providecommand{\noopsort}[1]{}\providecommand{\singleletter}[1]{#1}%

\end{document}